\documentclass[pdflatex,sn-nature]{sn-jnl}

\usepackage{graphicx}%
\usepackage{multirow}%
\usepackage{amsmath,amssymb,amsfonts}%
\usepackage{amsthm}%
\usepackage{mathrsfs}%
\usepackage[title]{appendix}%
\usepackage{xcolor}%
\usepackage{textcomp}%
\usepackage{manyfoot}%
\usepackage{booktabs}%
\usepackage{algorithm}%
\usepackage{algorithmicx}%
\usepackage{algpseudocode}%
\usepackage{listings}%

\usepackage[T1]{fontenc}

\begin{document}

\title[Article Title]{Observation of Quantum fluids of light in ``einstein'' aperiodic monotile}

\title[Article Title]{Realization of the "einstein" monotile with macroscopic coherent states}

\title[Article Title]{Polariton condensates meet an aperiodic ``einstein'' monotile}

\title[Article Title]{Observation of an aperiodic polariton monotile}


\author*[1]{\fnm{Sergey} \sur{Alyatkin}}\email{S.Alyatkin@skoltech.ru}

\author[1,2]{\fnm{Yaroslav V.} \sur{Kartashov}}

\author[1]{\fnm{Kirill} \sur{Sitnik}}

\author[1]{\fnm{Philipp} \sur{Grigoryev}}

\author*[1]{\fnm{Pavlos G.} \sur{Lagoudakis}}\email{P.Lagoudakis@skoltech.ru}

\affil[1]{\orgname{Hybrid Photonics Laboratory, Skolkovo Institute of Science and Technology}, \orgaddress{\street{Territory of Innovation Center Skolkovo, Bolshoy Boulevard 30, building 1}, \postcode{121205} \city{Moscow}, \country{Russia}}}

\affil[2]{\orgname{Institute of Spectroscopy of Russian Academy of Sciences}, \orgaddress{\street{Fizicheskaya Str., 5}, \postcode{108840} \city{Moscow}, \country{Russia}}}


\abstract{
A plethora of unconventional localization phenomena~\cite{Segev2013, Freedman_Nature2006, Wang2019, Wang_NatPho2024} and fractal features of linear spectrum~\cite{Bandres_PRX2016, Xu_NatPho2021} observed in quasiperiodic structures have been accompanied by a long-standing quest for the geometrical elements and structures that permit tilings of the plane, but only in a non-periodic manner. Until 2024~\cite{Smith_2024}, it was believed that such quasiperiodic structures, or quasicrystals, could only be composed of at least two different tiles. Surprisingly, a newly discovered class of quasicrystals requires only one elementary \textit{monotile}~\cite{Smith_2024}. However, its physical realization and study of propagating coherent excitations in this novel setting remained elusive. Here we optically sculpt aperiodic quasicrystals composed of ``einstein'' monotiles in an inorganic microcavity and observe nontrivial relative phases of the exciton-polariton condensates nonresonantly excited at the vertices of each monotile. Utilizing energy-resolved tomography in momentum-space, we reveal the formation of distinct Bragg peaks with six-fold symmetry and Dirac-like spectral fingerprints, intrinsic to the underlying graphene-like structure~\cite{Monotile_PRL}, while interferometric phase reconstruction shows a nontrivial synchronization pattern distinct from both periodic triangular lattices and Penrose quasicrystals. Our work demonstrates that monotiles can be converted into a programmable driven-dissipative artificial material, where long-range coherence coexists with enforced geometric aperiodicity, producing synchronization and spectral responses distinct from both periodic and conventional quasicrystalline tilings. 
}


%

\maketitle

Since the discovery of quasicrystals~\cite{Schechtman_1984PRL}, scientists have experimentally demonstrated numerous quasicrystalline structures and examined their properties using a variety of physical platforms including photonic~\cite{Vardeny_NatPho2013}, electronic~\cite{Collins_NatComm2017, Kempkes_NatPhys2019}, acoustic~\cite{Han2025} and plasmonic systems~\cite{Arjas2024}, cold atoms~\cite{Schreiber_Science2015, Viebahn_PRL2019} and structured semiconductor cavities~\cite{Tanese_PRL2014, Baboux_PRB2017, Goblot_NatPhy2020}. The unique structural and symmetry properties of quasicrystals, often unattainable in other natural or artificial materials, give rise to a wide range of novel physical phenomena and evolution scenarios that can be observed in these complex aperiodic materials. 

\begin{figure}
\centering
\includegraphics[width=\columnwidth]{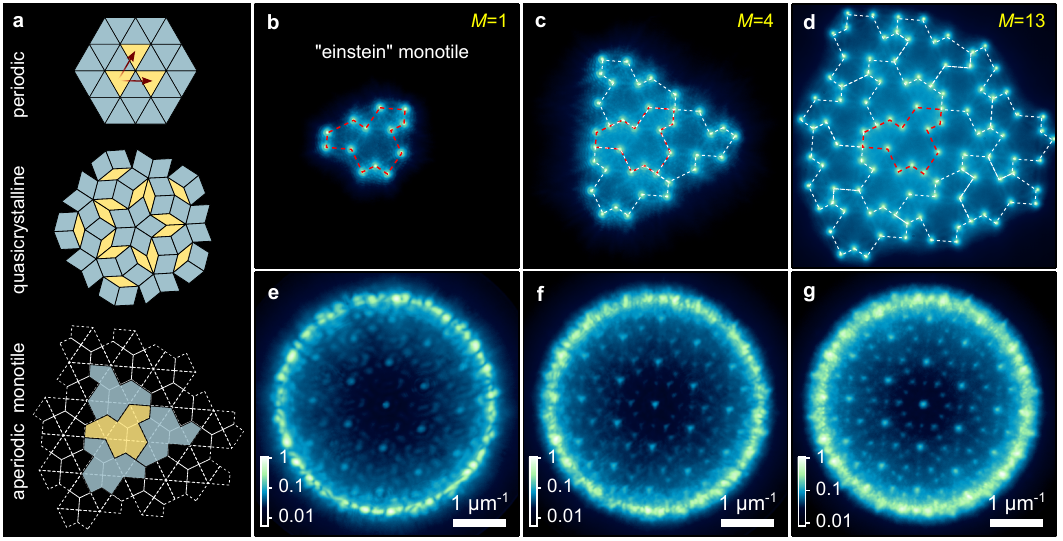}
\caption{\textbf{Build-up of an aperiodic 2D tiling based on the ``einstein'' (or hat) monotile using microcavity polaritons.} \textbf{a} Fragments of a periodic triangular lattice, Penrose quasicrystal, and an aperiodic tiling constructed with the ``einstein'' hat monotiles. \textbf{b} Experimental time-integrated real-space polariton photoluminescence (PL), excited by nonresonant structured pulsed laser pump in the shape of a single ($M=1$) ``einstein'' monotile. \textbf{c},\textbf{d} Show real-space polariton PL for the increased number of monotiles, $M=4$ and $M=13$, respectively. Overlaid dashed lines are to guide the eye and denote the position of the vertices with respect to the underlying  lattice. $\textbf{e}$,$\textbf{f}$,\textbf{g} Corresponding measured reciprocal-space polariton PL above condensation threshold with clearly visible Bragg peaks with $C_6$ discrete rotational symmetry.}\label{fig1}
\end{figure}

In contrast to periodic two-dimensional (2D) lattices, the quasicrystals do not possess translational symmetry (see Fig.~\ref{fig1}\textbf{a}), yet they feature the long-range order and, in some cases, discrete rotational symmetry that cannot be realized in periodic media~\cite{janot2012}. This is manifested, in particular, in fascinating fractal patterns of the Bragg peaks in far field diffraction patterns~\cite{Notomi_PRL2004, Vitiello_NatComm2014}. Self-similarity of aperiodic structures without periodicity results in fundamentally different evolution of excitations and peculiarities in transport properties. The reported studies on quasiperiodic systems have primarily focused on the propagation of waves~\cite{Matsui_Nature2007}, unconventional light localization in linear and nonlinear photonic systems of diverse rotational symmetries~\cite{Freedman_Nature2006, Xavier2010, Wang2019, Wang_NatPho2024}, and the impact of disorder on transport properties in quasicrystals~\cite{Levi2011}. Topological phenomena can also occur in quasiperiodic systems, since topology inevitably emerges in these structures as a direct consequence of system's dimension~\cite{Kraus_NatPhy2016, Bandres_PRX2016}.

Up to the seminal work by Smith \textit{et al.}~\cite{Smith_2024}, perhaps the most prominent and illustrative example of 2D aperiodic tiling with the smallest number of different building blocks was the Penrose mosaic, consisting of a thin and thick rhombuses~\cite{penrose1974role}, see the middle inset of Fig.~\ref{fig1}\textbf{a}. However, as proven by Smith and co-workers, the tessellation of the entire plane without any gaps can also be achieved by the discovered aperiodic ``einstein'' or hat \textit{monotile} (from the German “ein stein”, meaning “one stone”). A fragment of this aperiodic tiling is shown in the bottom inset of Fig.~\ref{fig1}\textbf{a}. The tiling rules dictate the coordinates of the vertices of this unique element, with some positioned at the vertices of an underlying hexagonal lattice used as a basis for monotile construction. As a result, theory predicts six-fold symmetry and Dirac-like cones in the momentum-resolved spectral function of the system~\cite{Monotile_PRL}. Among the unique properties of the discovered aperiodic monotile structure is chirality, which arises from the broken axial symmetry of the monotiles~\cite{Smith_2024_chiral, Moritake_arxiv2025}. The mathematical discovery of the aperiodic monotile thundered across multiple scientific fields ranging from geometry and computational physics to material science, optics, and condensed matter. However, the experimental implementation of quasicrystal structures (based on the aperiodic monotile) with tailored interactions between the lattice nodes, along with the investigation of their spectral properties, has remained challenging to date.

Here, we experimentally realize an all-optical 2D aperiodic monotile structure and demonstrate nonequilibrium condensation of microcavity exciton-polaritons in extended lattices. We achieve this by employing structured optical pumping, which creates ballistically propagating polaritons that predominantly condense at the vertices of the monotiles. The inorganic GaAs microcavity is nonresonantly excited above the condensation threshold with pulsed Ti:Sapphire laser. The pump profile is controlled with a programmable spatial light modulator (SLM), which transforms the initially non-structured laser beam into a desired aperiodic arrangement of diffracted beams, further focused on the sample. When the pump power exceeds a critical value, we observe the formation of a macroscopic coherent state with well-defined energy and phase. Using interferometry, we provide insights on the structure and physics of quantum fluids of light in a newly discovered aperiodic ``einstein'' tiling, on a par with other periodic and quasicrystalline arrays of polariton condensates.

First, we explore the spectral features of the aperiodic monotile structure and provide conclusive evidence of the phase locking between the condensates through analysis of the reciprocal-space polariton photoluminescence (PL) for the building blocks of the aperiodic structure. For this purpose, we optically imprint onto the microcavity the structures consisting of a variable number of the ``einstein'' monotiles $M=1, 4, 13$. Figure~\ref{fig1} shows measured real-space (top row) and corresponding reciprocal-space polariton PL (bottom row) above the condensation threshold, at $P\approx1.15P_\mathrm{thr}$. One can see that even for a single polariton monotile, shown in Fig.~\ref{fig1}\textbf{b}, corresponding reciprocal-space PL in Fig.~\ref{fig1}\textbf{e} reveals the formation of the narrow Bragg peaks, corresponding to six-fold symmetry of the underlying hexagonal lattice. Such a lattice and example monotile building blocks constructed on its basis are schematically depicted in the bottom inset in Fig.~\ref{fig1}\textbf{a}. We note that the number of well-resolvable distinct peaks in reciprocal space increases with the number of imprinted monotiles $M$ (compare  Fig.~\ref{fig1}\textbf{e} for $M=1$ and Fig.~\ref{fig1}\textbf{f} for $M=4$). This is further confirmed by experimental reciprocal-space PL distribution in Fig.~\ref{fig1}\textbf{g}, recorded for even larger structure with $M=13$, shown in Fig.~\ref{fig1}\textbf{d}. Superimposed with polariton PL white dashed contours in Fig.~\ref{fig1}\textbf{d} emphasize that pump laser spots are precisely positioned at the vertices of the interconnected hat monotiles.

\begin{figure}[t]
\centering
\includegraphics[width=\columnwidth]{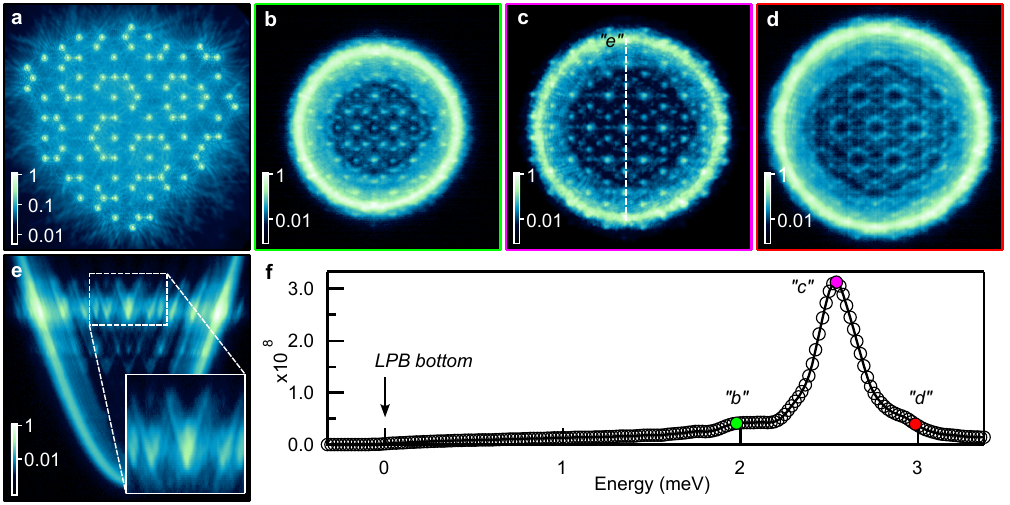}
\caption{\textbf{Energy tomography of 2D polariton quasicrystals based on the ``einstein'' tiling for $M=13$.} \textbf{a} Experimental time-integrated real-space polariton PL, excited by nonresonant pump. \textbf{b},\textbf{c},\textbf{d} Experimentally measured “slices” of energy paraboloid in reciprocal space [in $(k_x,k_y)$ plane] at energies $E=1.97$, $2.49$, and $2.94$ meV with respect to the bottom of the lower polariton branch. \textbf{e} Energy-resolved reciprocal space PL for $k_x=0$, schematically marked on \textbf{c} with vertical dashed line. The inset shows near perfectly linear dispersion. \textbf{f} Integrated total PL intensity in reciprocal space as a function of energy.}  \label{fig2}
\end{figure}

According to recent theoretical prediction~\cite{Monotile_PRL} the structure composed of the hat monotiles displays strong spectral similarities to that of graphene. In order to verify this experimentally, we realize well-established energy tomography technique~\cite{Alyatkin_NatComm} for the extended aperiodic lattice constructed of $M=13$ monotiles shown in Fig.~\ref{fig2}\textbf{a}. Utilizing measurements of the energy-resolved reciprocal-space PL for different $k_x$ cross-sections we reconstruct a three-dimensional energy paraboloid. Exemplary cross-sections of the latter at different energy levels are depicted in Fig.~\ref{fig2}\textbf{b},\textbf{c},\textbf{d} for $E=1.97, 2.49, 2.94$ meV. A fragment of the paraboloid reconstructed from measurements is shown in Supplementary Animation. We note that such reconstructed energy spectrum of the polariton monotile structure reveals remnants of Dirac-like cones (of course, the structure is aperiodic and does not possess a band-gap spectrum in the usual sense, but the fact that it is composed of monotiles on the basis of hexagonal lattice is immediately reflected in its energy spectrum). Here we refer the reader to iso-energy planes in reciprocal-space PL from higher to lower energies, Fig.~\ref{fig2}\textbf{d}-\textbf{b}, in which the characteristic transformation of the spectrum (upon passage through the energy level corresponding to the approximate position of Dirac points) is visible. Figure~\ref{fig2}\textbf{c} depicts bright Bragg peaks of expected six-fold symmetry. To further visualize the Dirac-like dispersion of polaritons, we extract the cross-section along white dashed line in Fig.~\ref{fig2}\textbf{c} and plot the result in Fig.~\ref{fig2}\textbf{e}. The inset with zoomed in energy-resolved reciprocal-space PL shows near perfectly linear dispersion with bright Dirac points. This observation highlights that the properties of polaritons in aperiodic tiling are inherited from hexagonal geometry of the underlying lattice.

\begin{figure}
\centering
\includegraphics[width=\columnwidth]{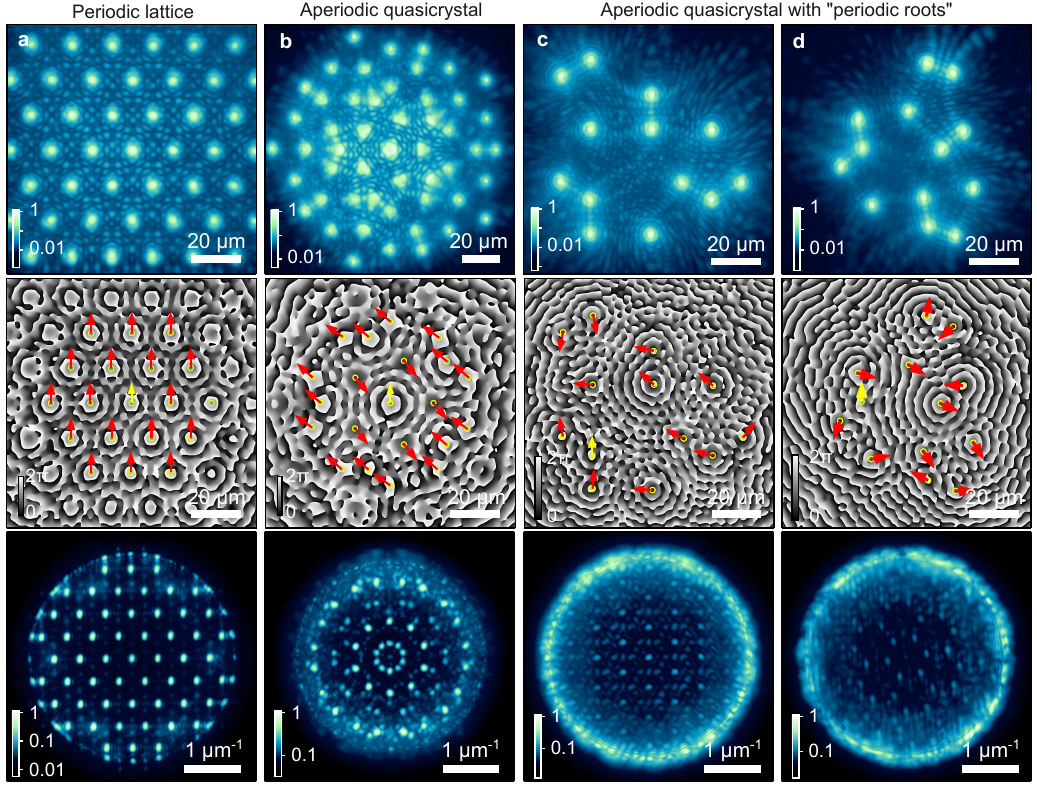}
\caption{\textbf{Phase locking of polariton condensates in periodic and quasicrystalline structures.} \textbf{a},\textbf{b},\textbf{c},\textbf{d} Experimental time-integrated real-space (top row) and reciprocal-space (bottom row) polariton PL above condensation threshold measured under CW excitation for the triangular lattice, the Penrose mosaic and the ``einstein'' monotiles for two different lattice constants of the underlying hexagonal lattice favouring either ``in-phase'' or ``out-of-phase'' synchronization between the nearest-neighbour condensates. Corresponding experimentally retrieved phase maps are shown in the middle row. Polaritons synchronize in-phase in periodic structure (\textbf{a}), yet adopt nontrivial phases in aperiodic structures (\textbf{b},\textbf{c},\textbf{d}).}\label{fig3}
\end{figure}

Cross-section of the energy paraboloid in Fig.~\ref{fig2}\textbf{e} reveals the tendency of polaritons to occupy the state with the certain energy, corresponding to the slice shown in Fig.~\ref{fig2}\textbf{c}. To confirm this, we integrate the polariton PL intensity for each energy-resolved slice and plot it as a function of energy. The obtained spectrum is presented in Fig.~\ref{fig2}\textbf{f}, where the coloured points (green, magenta and red) correspond to Fig.~\ref{fig2}\textbf{b},\textbf{c},\textbf{d}, respectively. We thus conclude that above the condensation threshold, polaritons indeed condense predominantly in a single energy state, characterized by six-fold symmetry in momentum space, as predicted in~\cite{Monotile_PRL}. However, in agreement with previous observations in lattices of ballistically propagating polaritons~\cite{Alyatkin_NatComm}, the overall structure of the energy paraboloid is rather complex (see Supplementary Animation). 

The inheritance by aperiodic lattices of certain spectral features from periodic hexagonal lattices raises fundamental questions, in particular, whether there is any difference between polaritons condensed in periodic versus aperiodic lattices, how their condensation dynamics differs between strictly periodic lattices and aperiodic ones, what distinguishes condensation in quasicrystals composed of, for example, two basic elements (like rhombi in the Penrose tiling) and the ``einstein'' monotile composed from one element, but with complex internal structure? To address these questions, we further explore the phases of the condensates positioned at the vertices of a regular triangular lattice, a Penrose tiling and the ``einstein'' monotile structure for pump amplitudes slightly above the condensation threshold. Under these conditions, the polariton lattices reach stationary and stable configurations corresponding to certain invariable arrangements of the condensate phases. To extract the relative phases of condensates, we utilize cw nonresonant excitation for the above-mentioned structures and apply the homodyne interferometric technique described elsewhere~\cite{Alyatkin_PRL,Alyatkin2024}.

Figure~\ref{fig3}\textbf{a} shows the real-space polariton PL and the measured phase map for a periodic triangular lattice. The chosen lattice constant $D=16.2$~$\mu$m facilitates the formation of synchronized in-phase condensates, as schematically depicted with red arrows (or ``phasors'') at different density maxima in real-space PL. We stress that, due to ballistic expansion of polaritons from the pumping spots, the phase varies away from the center of each spot. However, for this chosen spacing $D$, the coupling between condensates leads to stringent locking of their relative phases (in-phase configuration). In contrast, the Penrose quasicrystal in Fig.~\ref{fig3}\textbf{b} (with characteristic spacing between the central spots $D=17.1$~$\mu$m) exhibits a nontrivial phase difference of the condensates in the vertices of the aperiodic mosaic relative to the central one. Notably, due to the five‑fold rotational symmetry of the structure, condensates located at the same radial distance from the center (i.e., forming a single layer or circumference) tend to synchronize in phase. However, as we move on further from the center, we find the next layer to be counter-aligned.

Importantly, we found that aperiodic polariton arrays based on the monotiles exhibit a qualitatively different arrangement of phases in stable configurations compared to those observed in periodic lattices and the Penrose tiling. To illustrate this, we measured the relative phases of polariton condensates in the ``einstein'' hat monotile ($M=1$) for two different characteristic sizes, determined by the lattice constant $D$ of the underlying hexagonal lattice. Remarkably, for the monotile with $D=27.2$~$\mu$m, depicted in Fig.~\ref{fig3}\textbf{c}, we observe constructive interference between the nearest-neighbour condensates, indicating that they are in-phase. This predominant in-phase synchronization of the neighbouring condensates is also confirmed by the measured reciprocal-space PL. However, we note that some nodes with larger separation distances deviate from this common tendency, and their relative phase takes nontrivial values different from $0$ or $\pi$. We attribute this to the complex shape of the monotile, where, as can be seen from schematics in Fig.~\ref{fig1}\textbf{a} and \ref{fig1}\textbf{b}, neighboring condensates whose coupling is significant, can be separated by more than just two distinct distances (as opposed to the case of Penrose tiling), resulting in the presence of different coupling strengths (which in this dissipative system may be complex), and therefore in different possible phase arrangements. Moreover, in contrast to Penrose quasicrystals, the "einstein" monotile does not possess discrete rotational symmetry, and thus one cannot expect phase arrangements with equal phases on a ring of certain radius, akin to those shown in Fig.~\ref{fig3}\textbf{b}. As a result, the stable configuration emerging in the case of monotile for this value of $D$ is characterized by zero phase difference only for the \textit{nearest} condensates, but by a complex arrangement of phases between more distant condensates dictated by particular couplings. 

To further confirm this behavior, we decrease the monotile dimensions to $D=22.6$~$\mu$m and track the resulting change in the phase structure of emerging stable pattern. Corresponding images with experimentally measured real-space polariton PL above condensation threshold are presented in Fig.~\ref{fig3}\textbf{d}. In contrast to the larger monotile, the nearest condensates for this set $D$ value reveal predominant out-of-phase synchronization, as supported by the reciprocal-space PL and the retrieved phase map. At the same time, similar to observations for the larger structure, more distant condensates exhibit nontrivial mutual phases, determined by corresponding coupling strengths. These experimental results are fully confirmed by the results of numerical modeling of polariton condensation under nonresonant pump in aperiodic tilings with a different number of monotiles, see \textbf{Methods}. For any number of monotiles $M$, we find that the nearest condensates lock in-phase or out-of-phase, depending solely on the the lattice constant of the underlying hexagonal lattice, while condensates with larger spacing in the resulting pattern feature different phases. Surprisingly, the net phase pattern in aperiodic monotile structures can be decomposed into two sublattices of synchronized nodes, as shown in Extended Data Fig.~A1. The first sublattice consists of the nearest-neighbour condensates (denoted with red colour), while the second sublattice consists of the remaining, more distant (isolated) condensates, denoted in green. We note that both sublattices coexist, highlighting macroscopic coherence of the extended polariton system. This observation emphasizes the unique structure of the aperiodic monotile and clearly distinguishes it from Penrose quasicrystals and strictly periodic lattices. 
Therefore, from the phase measurements and theoretical analysis, we convincingly conclude that aperiodicity and absence of discrete rotational symmetry in the driven-dissipative polariton monotile system result in complex phase locking across the lattice. 

All in all, we have realized a 2D polariton aperiodic structure based on the recently discovered ``einstein'' monotile, using nonresonant structured excitation in an inorganic planar microcavity. We demonstrate that the absence of periodicity in the excitation pattern does not preclude phase locking of polariton condensates positioned at the vertices of the quasicrystal. The formation of macroscopic coherence above critical pump power is evidenced by narrow and intense Bragg peaks in reciprocal space. Phase-resolved interferometry reveals a nontrivial phase distribution across the tiling, similar to the observations in the Penrose quasicrystals~\cite{Alyatkin_Penrose}, and opposed to strictly periodic lattices, where all the nodes are strictly in-phase or out-of-phase~\cite{Toepfer_time-delay,Alyatkin_APL2024}. We emphasize that investigated aperiodic monotile structure has periodic origins rooted in the tessellation principle, which is intrinsically connected with an underlying hexagonal lattice. This results in observed Dirac-like dispersion of polaritons in the "einstein" monotile system.

Our results represent a step towards exploring many‑body physics and synchronization phenomena in aperiodic, nonlinear environment, with potential relevance to topological states. Future research should examine the role of chirality of the realized system. This would enable a more precise distinction between the newly discovered aperiodic monotile from, for example, famous Penrose quasicrystal, which lacks chiral properties. By utilizing the time-periodic excitation technique, recently demonstrated with polaritons~\cite{Gnusov_SciAdv,delValle_2023,delValle_2024}, one can also study shaken aperiodic lattices and assess the robustness of quasicrystalline order under controllable perturbations as function of potential depth and system symmetry.   

\section*{Methods}

\subsection*{Experiment}
The experimental measurements were carried out on a strain-compensated planar inorganic microcavity. The sample was cooled down to 4 K using a closed-cycle helium cryostat. The exciton-photon detuning was set to a negative value of $\delta=-4$ meV. This makes the polariton effective mass smaller and assists in node-to-node ballistic coupling~\cite{Toepfer_Optica, Alyatkin_PRL}. The non-resonant laser was tuned at the first Bragg minimum of the reflectivity stop-band (1.5578 eV) to improve excitation efficiency. The pulsed laser radiation ($\tau\approx5$ ps) at a fundamental repetition frequency of $\approx80$ MHz was additionally chopped using an acousto-optical modulator at a frequency of 5 kHz with a duty cycle of 3$\%$ to realize pulse train excitation. This ensured a stable set temperature of the sample even for a large number of vertices in the tiling. Such excitation regime was used in all time-integrated experiments with pulsed laser excitation for real- and reciprocal-space PL measurements (Fig.~\ref{fig1} and Fig.~\ref{fig2}). 

The results shown in Fig.~\ref{fig3} (top and bottom rows) were obtained with cw laser excitation modulated at a frequency of 5 kHz with a duty cycle of 1$\%$. The reconstructed phase maps in Fig.~\ref{fig3} were measured using interferometric techniques in a single-shot excitation regime with 50 $\mu$s pulses. 

\subsection*{Numerical modeling}
For numerical modeling of condensate evolution under nonresonant pumping, we used the 2D Gross-Pitaevskii equation for the polariton wavefunction $\Psi(\textbf{r},t)$ coupled to a rate equation for the exciton reservoir feeding the condensate $n(\textbf{r},t)$~\cite{Wouters_PRL2007}:

\begin{eqnarray}
 &&
\nonumber
 		i \hbar \frac{\partial \Psi}{\partial t} = -\frac{\hbar^2 }{2m_\mathrm{eff}} \nabla^2\Psi- i \frac{\hbar}{2} (\gamma_{c}-Rn)\Psi + g_{c} |\Psi|^2 \Psi+ g_{r}n  \Psi, \label{gpe1} 
 		\\ &&
		\frac{\partial n}{\partial t} =  -(\gamma_{r}+R|\Psi|^2)n + P(\textbf{r}). \label{gpe2}
\end{eqnarray}
Here $m_\mathrm{eff} \approx 5.63\times10^{-5}m_e$ is the effective mass for polaritons from the lower branch, where $m_e$ is the free electron rest mass, $g_{c}=2.4~\mu \textrm{eV}\mu\textrm{m}^2$ is the polariton-polariton and $g_{r}=2 g_c$ is the polariton-reservoir interaction strengths typical for GaAs-based microcavities, $R=0.021~\mu \textrm{m}^2\textrm{ps}^{-1}$ is the stimulated scattering rate of polaritons from reservoir, $\gamma_{c}\approx 0.182~\textrm{ps}^{-1}$ and $\gamma_{r}\approx 0.05~\textrm{ps}^{-1}$, are the decay rates for polariton condensate and reservoir excitons, respectively. The function $P(\textbf{r})=(P_0\gamma_{r}\gamma_{c}/R)\sum_{n}Q(\textbf{r}-\textbf{r}_{n})$ describes spatial profile of the pump consisting of identical Gaussian spots $Q(\textbf{r})=e^{-r^2/d^2}$ of width $d=1.5/(\textrm{ln}2)^{1/2}~\mu \textrm{m}$ (corresponding to individual pump spots with FWHM of $3~\mu\textrm{m}$ used in the experiment) that were placed in the vertices with coordinates $\textbf{r}_{n}$ of various aperiodic configurations constructed from $M=1$ to $M=13$ monotiles. The coordinates $\textbf{r}_{n}$ depend on the constant $D$ of the underlying hexagonal lattice that serves as a basis for construction of the monotile. We introduced here dimensionless pump amplitude $P_0$, allowing to conveniently determine threshold for condensation in inhomogeneous pump landscape $P(\textbf{r})$ and the domains of stability of the emerging aperiodic condensate lattices in units of $\gamma_{r}\gamma_{c}/R$, corresponding to condensation threshold for spatially uniform pump.

To determine all possible stationary states that can emerge in this system, the modeling of Eqs. (\ref{gpe1}) and (\ref{gpe2}) was performed using multiple realizations of small-scale noise for $\Psi|_{t=0}$ and $n|_{t=0}$, for each number of monotiles $M$ in configuration and for each pump amplitude $P_0$. To account for considerable ballistic expansion of polaritons from the pumped regions leading to nonzero polariton density even far from them, we used in modeling sufficiently large spatial domain $400\times400 ~\mu\textrm{m}^2$ greatly exceeding the pumped area. Numerical simulations show that above the condensation threshold $P_\textrm{thr}$, that decreases with increase of $M$, there exist the extended range of pump amplitudes $P_0$ where the excitation of stationary aperiodic polariton lattice occurs and the amplitudes of $|\Psi|$ and $n$ remain unchanged once the steady state is reached, while phase differences between neighboring condensates acquire constant values depending only on the constant $D$ of the underlying hexagonal lattice. There exist the alternating intervals of $D$, where nearest condensates are always in-phase or always out-of-phase, while the phases of condensates with larger spacing assume various values (that do not change with time $t$) depending on the particular configuration considered, as in experiments. Thus, for $D=27.2$~$\mu$m we encountered in-phase arrangement of the nearest-neighbour condensates, while for $D=22.6$~$\mu$m it switched to out-of-phase arrangement. The transition between the regimes, where only one of these configurations is stable, occurs within narrow range of $D$ values, where pumping typically results in irregular evolution of condensate and excitonic reservoir in time without the transition into steady-state regime.

Importantly, with increase of the number of monotiles $M$ in aperiodic structure the phase difference between closest condensates remains unchanged (across the entire structure) in comparison with the simplest $M=1$ case, for any value of $D$ indicating on the decisive role of such interactions even upon the formation of large patterns.

\backmatter

\bmhead{Supplementary information}

Supplementary information is available. 

\bmhead{Acknowledgements}
Authors acknowledge Dr. Helgi Sigur{\dh}sson for helpful discussion.

This study was supported by the Russian Science Foundation (RSF) (Grant No. 24-72-10118), https://rscf.ru/en/project/24-72-10118/.

\section*{Declarations}

\subsection*{Author contributions}
S.A. carried out the experiments and analyzed experimental data, K.S. and P.G. calculated the phase mask for the excitation profile, Y.V.K. performed analysis of the modes structure, S.A. and P.G.L. supervised the project, S.A. and Y.K. wrote draft of the manuscript, all authors contributed to discussion and writing the manuscript.  

\subsection*{Competing interests.} 

Authors declare that they have no competing interests.    

\subsection*{Data availability.} 
 
The data presented in this paper are available from the corresponding author upon reasonable request.

\subsection*{Additional information}
\textbf{Supplementary Information} 
Supplementary Animation with energy-resolved reciprocal space PL, corresponding to the structure shown in Fig.~\ref{fig3}.

\bigskip
\begin{flushleft}%
\bigskip\noindent
Springer journals and proceedings: \url{https://www.springer.com/gp/editorial-policies}

\bigskip\noindent
Nature Portfolio journals: \url{https://www.nature.com/nature-research/editorial-policies}

\end{flushleft}

\begin{appendices}

\section{Extended Data}\label{secA1}
\begin{figure}[t!]
\centering
\includegraphics[width=\columnwidth]{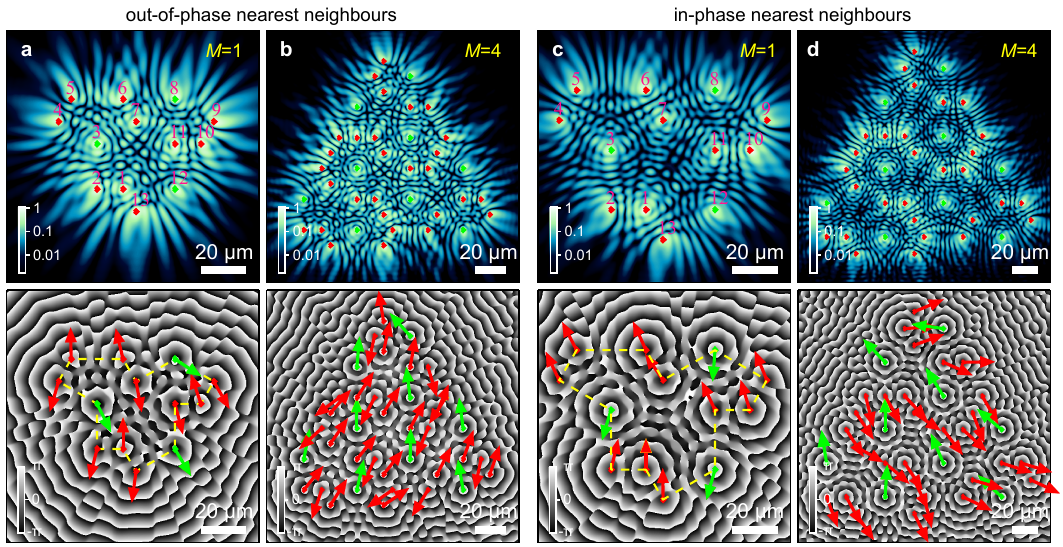}
\caption{\textbf{Phase locking of polariton condensates in aperiodic ``einstein'' monotile structures.} \textbf{a},\textbf{b},\textbf{c},\textbf{d} Simulated real-space densities and phases for the ``einstein'' monotiles of different lattice constant of the underlying hexagonal lattice, favouring either ``in-phase'' or ``out-of-phase'' coupling between the nearest condensates (denoted with red arrows), while more distant nodes (denoted with green arrows) form synchronized sublattice with different relative phases.}
\label{fig5}
\end{figure}

Extended Data Fig.~A1 shows densities and phases of numerically simulated eigenmodes of the aperiodic monotile structure for $M=1$ and $M=4$. 



\end{appendices}

\newpage
\bibliography{monotile}

@Misc{methods,
  note = {Materials and methods are available as supplementary material},
}

@article{Schechtman_1984PRL,
  title = {Metallic Phase with Long-Range Orientational Order and No Translational Symmetry},
  author = {Shechtman, D. and Blech, I. and Gratias, D. and Cahn, J. W.},
  journal = {Phys. Rev. Lett.},
  volume = {53},
  issue = {20},
  pages = {1951--1953},
  numpages = {0},
  year = {1984},
  month = {Nov},
  publisher = {American Physical Society},
  doi = {10.1103/PhysRevLett.53.1951},
  url = {https://link.aps.org/doi/10.1103/PhysRevLett.53.1951}
}

@Article{Vardeny_NatPho2013,
author={Vardeny, Z. Valy
and Nahata, Ajay
and Agrawal, Amit},
title={Optics of photonic quasicrystals},
journal={Nature Photonics},
year={2013},
month={Mar},
day={01},
volume={7},
number={3},
pages={177-187},
issn={1749-4893},
doi={10.1038/nphoton.2012.343},
url={https://doi.org/10.1038/nphoton.2012.343}
}

@article{Viebahn_PRL2019,
  title = {Matter-Wave Diffraction from a Quasicrystalline Optical Lattice},
  author = {Viebahn, Konrad and Sbroscia, Matteo and Carter, Edward and Yu, Jr-Chiun and Schneider, Ulrich},
  journal = {Phys. Rev. Lett.},
  volume = {122},
  issue = {11},
  pages = {110404},
  numpages = {6},
  year = {2019},
  month = {Mar},
  publisher = {American Physical Society},
  doi = {10.1103/PhysRevLett.122.110404},
  url = {https://link.aps.org/doi/10.1103/PhysRevLett.122.110404}
}

@Article{Xu_NatPho2021,
author={Xu, Xiao-Yun
and Wang, Xiao-Wei
and Chen, Dan-Yang
and Smith, C. Morais
and Jin, Xian-Min},
title={Quantum transport in fractal networks},
journal={Nature Photonics},
year={2021},
month={Sep},
day={01},
volume={15},
number={9},
pages={703-710},
issn={1749-4893},
doi={10.1038/s41566-021-00845-4},
url={https://doi.org/10.1038/s41566-021-00845-4}
}

@article{Tanese_PRL2014,
  title = {Fractal Energy Spectrum of a Polariton Gas in a Fibonacci Quasiperiodic Potential},
  author = {Tanese, D. and Gurevich, E. and Baboux, F. and Jacqmin, T. and Lema\^{\i}tre, A. and Galopin, E. and Sagnes, I. and Amo, A. and Bloch, J. and Akkermans, E.},
  journal = {Phys. Rev. Lett.},
  volume = {112},
  issue = {14},
  pages = {146404},
  numpages = {5},
  year = {2014},
  month = {Apr},
  publisher = {American Physical Society},
  doi = {10.1103/PhysRevLett.112.146404},
  url = {https://link.aps.org/doi/10.1103/PhysRevLett.112.146404}
}

@Article{Collins_NatComm2017,
author={Collins, Laura C.
and Witte, Thomas G.
and Silverman, Rochelle
and Green, David B.
and Gomes, Kenjiro K.},
title={Imaging quasiperiodic electronic states in a synthetic Penrose tiling},
journal={Nature Communications},
year={2017},
month={Jun},
day={22},
volume={8},
number={1},
pages={15961},
issn={2041-1723},
doi={10.1038/ncomms15961},
url={https://doi.org/10.1038/ncomms15961}
}

@Article{Matsui_Nature2007,
author={Matsui, Tatsunosuke
and Agrawal, Amit
and Nahata, Ajay
and Vardeny, Z. Valy},
title={Transmission resonances through aperiodic arrays of subwavelength apertures},
journal={Nature},
year={2007},
month={Mar},
day={01},
volume={446},
number={7135},
pages={517-521},
issn={1476-4687},
doi={10.1038/nature05620},
url={https://doi.org/10.1038/nature05620}
}

@article{penrose1974role,
  title={The role of aesthetics in pure and applied mathematical research},
  author={Penrose, Roger},
  journal={Bull. Inst. Math. Appl.},
  volume={10},
  pages={266--271},
  year={1974}
}

@article{Smith_2024,
  title = {An aperiodic monotile},
  volume = {4},
  ISSN = {2766-1334},
  url = {http://dx.doi.org/10.5070/C64163843},
  DOI = {10.5070/c64163843},
  number = {1},
  journal = {Combinatorial Theory},
  publisher = {California Digital Library (CDL)},
  author = {Smith,  David and Myers,  Joseph Samuel and Kaplan,  Craig S. and Goodman-Strauss,  Chaim},
  year = {2024},
  month = jul 
}

@article{Schreiber_Science2015,
author = {Michael Schreiber  and Sean S. Hodgman  and Pranjal Bordia  and Henrik P. Lüschen  and Mark H. Fischer  and Ronen Vosk  and Ehud Altman  and Ulrich Schneider  and Immanuel Bloch },
title = {Observation of many-body localization of interacting fermions in a quasirandom optical lattice},
journal = {Science},
volume = {349},
number = {6250},
pages = {842-845},
year = {2015},
doi = {10.1126/science.aaa7432},
URL = {https://www.science.org/doi/abs/10.1126/science.aaa7432},
eprint = {https://www.science.org/doi/pdf/10.1126/science.aaa7432}}

@Article{Kempkes_NatPhys2019,
author={Kempkes, S. N.
and Slot, M. R.
and Freeney, S. E.
and Zevenhuizen, S. J. M.
and Vanmaekelbergh, D.
and Swart, I.
and Smith, C. Morais},
title={Design and characterization of electrons in a fractal geometry},
journal={Nature Physics},
year={2019},
month={Feb},
day={01},
volume={15},
number={2},
pages={127-131},
issn={1745-2481},
doi={10.1038/s41567-018-0328-0},
url={https://doi.org/10.1038/s41567-018-0328-0}
}

@article{Alyatkin_APL2024,
    author = {Alyatkin, Sergey and Sigurðsson, Helgi and Kartashov, Yaroslav V. and Gnusov, Ivan and Sitnik, Kirill and Töpfer, Julian D. and Lagoudakis, Pavlos G.},
    title = "{All-optical triangular and honeycomb lattices of exciton–polaritons}",
    journal = {Applied Physics Letters},
    volume = {124},
    number = {6},
    pages = {062105},
    year = {2024},
    month = {02},
    issn = {0003-6951},
    doi = {10.1063/5.0180272},
    url = {https://doi.org/10.1063/5.0180272}
}

@Article{Freedman_Nature2006,
author={Freedman, Barak
and Bartal, Guy
and Segev, Mordechai
and Lifshitz, Ron
and Christodoulides, Demetrios N.
and Fleischer, Jason W.},
title={Wave and defect dynamics in nonlinear photonic quasicrystals},
journal={Nature},
year={2006},
month={Apr},
day={01},
volume={440},
number={7088},
pages={1166-1169},
issn={1476-4687},
doi={10.1038/nature04722},
url={https://doi.org/10.1038/nature04722}
}

@Article{Vitiello_NatComm2014,
author={Vitiello, Miriam Serena
and Nobile, Michele
and Ronzani, Alberto
and Tredicucci, Alessandro
and Castellano, Fabrizio
and Talora, Valerio
and Li, Lianhe
and Linfield, Edmund H.
and Davies, A. Giles},
title={Photonic quasi-crystal terahertz lasers},
journal={Nature Communications},
year={2014},
month={Dec},
day={19},
volume={5},
number={1},
pages={5884},
issn={2041-1723},
doi={10.1038/ncomms6884},
url={https://doi.org/10.1038/ncomms6884}
}

@article{Notomi_PRL2004,
  title = {Lasing Action due to the Two-Dimensional Quasiperiodicity of Photonic Quasicrystals with a Penrose Lattice},
  author = {Notomi, M. and Suzuki, H. and Tamamura, T. and Edagawa, K.},
  journal = {Phys. Rev. Lett.},
  volume = {92},
  issue = {12},
  pages = {123906},
  numpages = {4},
  year = {2004},
  month = {Mar},
  publisher = {American Physical Society},
  doi = {10.1103/PhysRevLett.92.123906},
  url = {https://link.aps.org/doi/10.1103/PhysRevLett.92.123906}
}

@article{Baboux_PRB2017,
  title = {Measuring topological invariants from generalized edge states in polaritonic quasicrystals},
  author = {Baboux, Florent and Levy, Eli and Lema\^{\i}tre, Aristide and G\'omez, Carmen and Galopin, Elisabeth and Le Gratiet, Luc and Sagnes, Isabelle and Amo, Alberto and Bloch, Jacqueline and Akkermans, Eric},
  journal = {Phys. Rev. B},
  volume = {95},
  issue = {16},
  pages = {161114},
  numpages = {5},
  year = {2017},
  month = {Apr},
  publisher = {American Physical Society},
  doi = {10.1103/PhysRevB.95.161114},
  url = {https://link.aps.org/doi/10.1103/PhysRevB.95.161114}
}

@Article{Kraus_NatPhy2016,
author={Kraus, Yaacov E.
and Zilberberg, Oded},
title={Quasiperiodicity and topology transcend dimensions},
journal={Nature Physics},
year={2016},
month={Jul},
day={01},
volume={12},
number={7},
pages={624-626},
issn={1745-2481},
doi={10.1038/nphys3784},
url={https://doi.org/10.1038/nphys3784}
}

@Article{Wang_NatPho2024,
author={Wang, Peng
and Fu, Qidong
and Konotop, Vladimir V.
and Kartashov, Yaroslav V.
and Ye, Fangwei},
title={Observation of localization of light in linear photonic quasicrystals with diverse rotational symmetries},
journal={Nature Photonics},
year={2024},
month={Jan},
day={04},
issn={1749-4893},
doi={10.1038/s41566-023-01350-6},
url={https://doi.org/10.1038/s41566-023-01350-6}
}

@book{janot2012,
  title     = "Quasicrystals: A Primer",
  author    = "Janot, C.",
  year      = 2012,
  publisher = "Clarendon Press",
  address   = "Oxford"
}

@Article{Goblot_NatPhy2020,
author={Goblot, V.
and {\v{S}}trkalj, A.
and Pernet, N.
and Lado, J. L.
and Dorow, C.
and Lema{\^i}tre, A.
and Le Gratiet, L.
and Harouri, A.
and Sagnes, I.
and Ravets, S.
and Amo, A.
and Bloch, J.
and Zilberberg, O.},
title={Emergence of criticality through a cascade of delocalization transitions in quasiperiodic chains},
journal={Nature Physics},
year={2020},
month={Aug},
day={01},
volume={16},
number={8},
pages={832-836},
issn={1745-2481},
doi={10.1038/s41567-020-0908-7},
url={https://doi.org/10.1038/s41567-020-0908-7}
}

@Article{Alyatkin_NatComm, 
author={Alyatkin, S.
and Sigurdsson, H.
and Askitopoulos, A.
and T{\"o}pfer, J. D.
and Lagoudakis, P. G.},
title={Quantum fluids of light in all-optical scatterer lattices},
journal={Nature Communications},
year={2021},
month={Sep},
day={22},
volume={12},
number={1},
pages={5571},
issn={2041-1723},
doi={10.1038/s41467-021-25845-4},
url={https://doi.org/10.1038/s41467-021-25845-4}
}

@article{Toepfer_Optica,
author = "J. D. T{\"{o}}pfer and I. Chatzopoulos and H. Sigurdsson and T. Cookson and Y. G. Rubo and P. G. Lagoudakis",
journal = "Optica",
number = "1",
pages = "106--113",
publisher = "Optica Publishing Group",
title = "Engineering spatial coherence in lattices of polariton condensates",
volume = "8",
month = Jan,
year = "2021",
url = "https://opg.optica.org/optica/abstract.cfm?URI=optica-8-1-106",
doi = "10.1364/OPTICA.409976",
}

@article{Alyatkin_PRL,
  title = {Optical Control of Couplings in Polariton Condensate Lattices},
  author = {Alyatkin, S. and T\"opfer, J. D. and Askitopoulos, A. and Sigurdsson, H. and Lagoudakis, P. G.},
  journal = {Phys. Rev. Lett.},
  volume = {124},
  issue = {20},
  pages = {207402},
  numpages = {6},
  year = {2020},
  month = {May},
  publisher = {American Physical Society},
  doi = {10.1103/PhysRevLett.124.207402},
  url = {https://link.aps.org/doi/10.1103/PhysRevLett.124.207402}
}

@article{Toepfer_time-delay,
author={T{\"o}pfer, J. D.
and Sigurdsson, H.
and Pickup, L.
and Lagoudakis, P. G.},
title={Time-delay polaritonics},
journal={Communications Physics},
year={2020},
month={Jan},
day={07},
volume={3},
number={1},
pages={2},
issn={2399-3650},
doi={10.1038/s42005-019-0271-0},
url={https://doi.org/10.1038/s42005-019-0271-0}
}

@article{Bandres_PRX2016,
  title = {Topological Photonic Quasicrystals: Fractal Topological Spectrum and Protected Transport},
  author = {Bandres, Miguel A. and Rechtsman, Mikael C. and Segev, Mordechai},
  journal = {Phys. Rev. X},
  volume = {6},
  issue = {1},
  pages = {011016},
  numpages = {12},
  year = {2016},
  month = {Feb},
  publisher = {American Physical Society},
  doi = {10.1103/PhysRevX.6.011016},
  url = {https://link.aps.org/doi/10.1103/PhysRevX.6.011016}
}

@article{Monotile_PRL,
  title = {Physical Properties of an Aperiodic Monotile with Graphene-like Features, Chirality, and Zero Modes},
  author = {Schirmann, Justin and Franca, Selma and Flicker, Felix and Grushin, Adolfo G.},
  journal = {Phys. Rev. Lett.},
  volume = {132},
  issue = {8},
  pages = {086402},
  numpages = {8},
  year = {2024},
  month = {Feb},
  publisher = {American Physical Society},
  doi = {10.1103/PhysRevLett.132.086402},
  url = {https://link.aps.org/doi/10.1103/PhysRevLett.132.086402}
}

@article{Wang2019,
  title = {Localization and delocalization of light in photonic moiré lattices},
  volume = {577},
  ISSN = {1476-4687},
  url = {http://dx.doi.org/10.1038/s41586-019-1851-6},
  DOI = {10.1038/s41586-019-1851-6},
  number = {7788},
  journal = {Nature},
  publisher = {Springer Science and Business Media LLC},
  author = {Wang,  Peng and Zheng,  Yuanlin and Chen,  Xianfeng and Huang,  Changming and Kartashov,  Yaroslav V. and Torner,  Lluis and Konotop,  Vladimir V. and Ye,  Fangwei},
  year = {2019},
  month = dec,
  pages = {42–46}
}

@article{Segev2013,
  title = {Anderson localization of light},
  volume = {7},
  ISSN = {1749-4893},
  url = {http://dx.doi.org/10.1038/NPHOTON.2013.30},
  DOI = {10.1038/nphoton.2013.30},
  number = {3},
  journal = {Nature Photonics},
  publisher = {Springer Science and Business Media LLC},
  author = {Segev,  Mordechai and Silberberg,  Yaron and Christodoulides,  Demetrios N.},
  year = {2013},
  month = feb,
  pages = {197–204}
}

@article{Alyatkin2024,
  title = {Antiferromagnetic Ising model in a triangular vortex lattice of quantum fluids of light},
  volume = {10},
  ISSN = {2375-2548},
  url = {http://dx.doi.org/10.1126/sciadv.adj1589},
  DOI = {10.1126/sciadv.adj1589},
  number = {34},
  journal = {Science Advances},
  publisher = {American Association for the Advancement of Science (AAAS)},
  author = {Alyatkin,  Sergey and Milián,  Carles and Kartashov,  Yaroslav V. and Sitnik,  Kirill A. and Gnusov,  Ivan and T\"{o}pfer,  Julian D. and Sigurðsson,  Helgi and Lagoudakis,  Pavlos G.},
  year = {2024},
  month = aug 
}

@article{Smith_2024_chiral,
  title = {A chiral aperiodic monotile},
  volume = {4},
  ISSN = {2766-1334},
  url = {http://dx.doi.org/10.5070/C64264241},
  DOI = {10.5070/c64264241},
  number = {2},
  journal = {Combinatorial Theory},
  publisher = {California Digital Library (CDL)},
  author = {Smith,  David and Myers,  Joseph Samuel and Kaplan,  Craig S. and Goodman-Strauss,  Chaim},
  year = {2024},
  month = sep 
}

@misc{Alyatkin_Penrose,
  doi = {10.48550/ARXIV.2409.16801},
  url = {https://arxiv.org/abs/2409.16801},
  author = {Alyatkin,  Sergey and Sitnik,  Kirill and Daníelsson,  Valtýr Kári and Kartashov,  Yaroslav V. and T\"{o}pfer,  Julian D. and Sigurðsson,  Helgi and Lagoudakis,  Pavlos G.},
  title = {Quantum Fluids of Light in 2D Artificial Reconfigurable Aperiodic Crystals with Tailored Coupling},
  publisher = {arXiv},
  year = {2024},
  copyright = {Creative Commons Attribution 4.0 International}
}

@article{Gnusov_SciAdv,
  title = {Quantum vortex formation in the “rotating bucket” experiment with polariton condensates},
  volume = {9},
  ISSN = {2375-2548},
  url = {http://dx.doi.org/10.1126/sciadv.add1299},
  DOI = {10.1126/sciadv.add1299},
  number = {4},
  journal = {Science Advances},
  publisher = {American Association for the Advancement of Science (AAAS)},
  author = {Gnusov,  Ivan and Harrison,  Stella and Alyatkin,  Sergey and Sitnik,  Kirill and T\"{o}pfer,  Julian and Sigurdsson,  Helgi and Lagoudakis,  Pavlos},
  year = {2023},
  month = jan 
}

@article{delValle_2023,
  title = {Optically Driven Rotation of Exciton–Polariton Condensates},
  volume = {23},
  ISSN = {1530-6992},
  url = {http://dx.doi.org/10.1021/acs.nanolett.3c01021},
  DOI = {10.1021/acs.nanolett.3c01021},
  number = {10},
  journal = {Nano Letters},
  publisher = {American Chemical Society (ACS)},
  author = {del Valle-Inclan Redondo,  Yago and Schneider,  Christian and Klembt,  Sebastian and H\"{o}fling,  Sven and Tarucha,  Seigo and Fraser,  Michael D.},
  year = {2023},
  month = may,
  pages = {4564–4571}
}

@article{delValle_2024,
  title = {Non-reciprocal band structures in an exciton–polariton Floquet optical lattice},
  volume = {18},
  ISSN = {1749-4893},
  url = {http://dx.doi.org/10.1038/s41566-024-01424-z},
  DOI = {10.1038/s41566-024-01424-z},
  number = {6},
  journal = {Nature Photonics},
  publisher = {Springer Science and Business Media LLC},
  author = {del Valle Inclan Redondo,  Yago and Xu,  Xingran and Liew,  Timothy C. H. and Ostrovskaya,  Elena A. and Stegmaier,  Alexander and Thomale,  Ronny and Schneider,  Christian and Dam,  Siddhartha and Klembt,  Sebastian and H\"{o}fling,  Sven and Tarucha,  Seigo and Fraser,  Michael D.},
  year = {2024},
  month = apr,
  pages = {548–553}
}

@misc{Moritake_arxiv2025,
  doi = {10.48550/ARXIV.2506.07561},
  url = {https://arxiv.org/abs/2506.07561},
  author = {Moritake,  Yuto and Takiguchi,  Masato and Aihara,  Takuma and Notomi,  Masaya},
  title = {Chiral Diffraction from Aperiodic Monotile Lattice},
  publisher = {arXiv},
  year = {2025},
  copyright = {Creative Commons Attribution Non Commercial No Derivatives 4.0 International}
}

@article{Arjas2024,
  title = {High topological charge lasing in quasicrystals},
  volume = {15},
  ISSN = {2041-1723},
  url = {http://dx.doi.org/10.1038/s41467-024-53952-5},
  DOI = {10.1038/s41467-024-53952-5},
  number = {1},
  journal = {Nature Communications},
  publisher = {Springer Science and Business Media LLC},
  author = {Arjas,  Kristian and Taskinen,  Jani Matti and Heilmann,  Rebecca and Salerno,  Grazia and T\"{o}rm\"{a},  P\"{a}ivi},
  year = {2024},
  month = nov 
}

@article{Han2025,
  title = {Observation of dispersive acoustic quasicrystals},
  volume = {16},
  ISSN = {2041-1723},
  url = {http://dx.doi.org/10.1038/s41467-025-57067-3},
  DOI = {10.1038/s41467-025-57067-3},
  number = {1},
  journal = {Nature Communications},
  publisher = {Springer Science and Business Media LLC},
  author = {Han,  Chenglin and Chen,  Li-Qun and Yang,  Tianzhi and Xu,  Guoqiang and Li,  Jiaxin and Li,  Changyou and Fan,  Haiyan and Alù,  Andrea and Qiu,  Cheng-Wei},
  year = {2025},
  month = feb 
}

@article{Xavier2010,
  title = {Reconfigurable optically induced quasicrystallographic three-dimensional complex nonlinear photonic lattice structures},
  volume = {22},
  ISSN = {2041-1723},
  number = {1},
  journal = {Advanced Materials},
  publisher = {Springer Science and Business Media LLC},
  author = {Xavier, J and Boguslawski, M and Rose, P and Joseph, J and Denz, C},
  year = {2010},
  pages = {356–360}
}

@article{levi2011,
  title = {Disorder-enhanced transport in photonic quasicrystals},
  volume = {332},
  number = {6037},
  journal = {Science},
  publisher = {Springer Science and Business Media LLC},
  author = {Levi, L and Rechtsman, M and Freedman, B and Schwartz, T and Manela, O and Segev, M},
  year = {2011},
  pages = {1541-1544}
  }

@article{Wouters_PRL2007,
  title = {Excitations in a Nonequilibrium {B}ose-{E}instein Condensate of Exciton Polaritons},
  author = {Wouters, Michiel and Carusotto, Iacopo},
  journal = {Phys. Rev. Lett.},
  volume = {99},
  issue = {14},
  pages = {140402},
  numpages = {4},
  year = {2007},
  month = {Oct},
  publisher = {American Physical Society},
  doi = {10.1103/PhysRevLett.99.140402},
  url = {https://link.aps.org/doi/10.1103/PhysRevLett.99.140402}
}

\end{document}